\documentclass[10pt,twocolumn]{article}
\usepackage{graphicx}
\usepackage{multicol}
\newcommand{\dc}{$^\circ$}

\begin{document}

\twocolumn[%
\begin{center}
{\bf The Effects of Temperature, Pressure, and Humidity Variations on 100~Meter Sprint Performances}\\
\vskip 1mm
J. R. Mureika \\
Department of Physics, Loyola Marymount University, Los Angeles, CA  90045 \\
Email: jmureika@lmu.edu~~~~Web:  http://myweb.lmu.edu/jmureika/
\end{center}
\vskip 2mm
]

\noindent{\bf INTRODUCTION}\\
It is well known that ``equivalent'' sprint race times run with different
accompanying wind speeds or at different altitudes are anything but equivalent
races.  
Adjusting these times for atmospheric drag effects has been
the focus of many past studies, including but not limited to [1,2,3,4,5]
and references therein.  The drag force acting on a sprinter running at
speed $v$ in a wind $w$ is a function of air density and the relative 
wind speed, 
$F_{\rm drag} \propto \rho_{\rm air} (v-w)^2,$
where density has traditionally been calculated using the race venue's 
elevation above sea level.\\

However, air density variation is dependent on more than just altitude.
This work will quantify how changes in air temperature, barometric
pressure, and humidity levels influence 100~m sprint performances.  These
variables collectively determine an effective altitude known as {\it density
altitude}, which depending on atmospheric conditions can be significantly
different than a venue's physical elevation above sea level.  The density
of hot air is low, yielding a higher density altitude and thus simulating
and increase in physical altitude.  Increased humidity levels have a similar
effect.  Conversely, colder temperatures and lower humidity can potentially
simulate a decrease in physical altitude.\\

\noindent{\bf METHODS}\\
The numerical model of sprinting performances used in
Reference [4] is modified using standard hydrodynamic principles to
include the effects of air temperature, pressure, and humidity levels on
aerodynamic drag.  Race times are obtained by numerically-integrating the
associated equations for various temperatures in the range 15-35\dc,
relative humidity levels (RH) between 0 and 100\%, and atmospheric pressures
between 85-105~kPa.  These calculations are performed for wind speeds
between -3~m/s and +3~m/s.  The resulting data are then 
compared to a defined standard race with no wind, sea 
level atmospheric pressure and temperature of 25\dc.\\

\noindent{\bf RESULTS AND DISCUSSION}\\
Temperature alone does not have a profound impact on the simulations,
giving a performance differential of 0.02~s over the 20\dc
range.   Wind and physical altitude corrections under these conditions are thus
essentially identical to those discussed in the literature, confirming
the earlier result of [3]. The
combined effect of relative humidity and temperature plays a slightly 
more crucial role in adjusting performances, since the air density of hot, 
saturated air is significantly lower than that of colder dry air.
\begin{figure}[h]
\includegraphics[scale=0.33,angle=270]{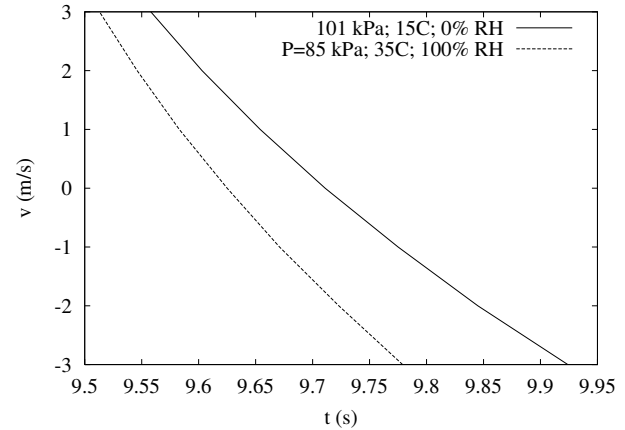}
\caption{\small Range of wind-aided and hindered (+3~to -3~m/s) performances 
equivalent to a ``standard'' 9.70~second 100~m sprint at sea-level with no wind for
extreme high-pressure (top curve) and low pressure (bottom curve) conditions.}
\label{Fig1}
\end{figure}

When the effects of pressure, temperature, and relative humidity 
changes are considered in combination, the corrections to performances can 
be very large.  
Figure~1 shows the range of possible simulation times, bounded
by times run in extreme conditions: 85~kPa, 100\% RH, and 35\dc~(yielding the
least dense atmosphere) and 105~kPa, 0\% RH, 15\dc~(most dense).
The race times vary by up to or over 0.1~seconds between 
these extremes even after wind correction is taken into account.
The results suggest that a non-negligible difference in race
times can be expected for ``equivalent'' performances run 
with the same wind speed at the same venue or physical altitude, but 
under different atmospheric conditions.\\

\noindent{\bf REFERENCES}\\
\noindent[1]  C.\ T.\ M.\ Davies, {\it J.\ Appl.\ Physio.} {\bf 48}, 702-709
(1980)\\
\noindent[2] Ward-Smith, A.\ J\., {\it J. Biomech.} {\bf 17}, 339-347 (1984); {\it J.  Biomech.} {\bf 18}, 351-357 (1985); {\it J. Sport. Sci.} {\bf 17}, 325-334 (1999)\\
\noindent[3] Dapena, J. and Feltner, M. E., {\it Int. J. Sport Biomech.} {\bf 3}, 
6-39 (1987); Dapena, J., in {\it The Big Green Book}, Track and Field News Press (2000)\\
\noindent[4] Linthorne, N. P., {\it J. App. Biomech.} {\bf 10}, 110-131 (1994)\\
\noindent[5] Mureika, J. R., {\it Can. J. Phys.} {\bf 79}, 697-713 (2001); {\it New Stud. Athl.} {\bf 15 (3/4)}, 53-58 (2000); {\it Can. J. Phys.} {\bf 81}, 895-910 (2003)\\

\noindent{\bf ACKNOWLEDGMENTS}\\
This work is made possible through the generous financial support of Loyola
Marymount University.
 
\end{document}